\documentclass[10pt]{iopart}
\usepackage{graphicx}
\begin{document}

\title[]{Understanding of the Vortex Lattice Behavior Induced by Selective Removal of Atoms from a Bose-Einstein Condensate}

\author{Hui Zhai and Qi Zhou}

\address{Center for Advanced Study, Tsinghua University, Beijing, China}

\begin{abstract}

It is known that the quantized vortices in a superfluid can be
described by a dual electromagnetic model through the duality
transformation. Recently a new technique, which can selectively
remove atoms from a Bose-Einstein condensate, was applied to the
vortex lattice state of the rapid rotating Boson gases. The
increase of vortex number, dynamical formation of giant vortex,
the oscillation and damping of vortex motion were observed in
these experiments. In this paper we will discuss these
observations in the framework of dual description and show how to
understood these observations naturally.

\end{abstract}



\maketitle

\section{Introduction}

It is well known that quantized vortices play an important role in
understanding many phenomena of superfluid, superconductivity and
even astrophysics. In the latest several years a remarkable
experimental development has been made in the rapidly rotating
Bose-Einstein condensate(BEC). A large amount of angular momentum
is imparted into the system and its rotation frequency becomes
faster and faster. The ground state of the fast rotating BEC was
observed to be a regular triangular vortex lattice at nearly zero
temperature. This provides a new stage of studying both the
equilibrium and the dynamic properties of the vortex matter.

Recently two novel experiments were reported by Cornell's group in
JILA.\cite{cornell}\cite{cornellsecond} Atoms in the hyperfine
spin $|F=1\rangle$ state, were rotated rapidly by the evaporative
spin-up mechanism. About $180$ vortices were observed in the cloud
and they were arranged into a regular lattice structure. And then,
an atom removal laser, whose frequency was tuned to the transition
energy between the $|F=1\rangle$ state and the $|F=0\rangle$
state, was applied to the center of the condensate, and the recoil
from a spontaneously scatted photon blasts atoms out of the
condensate. In this way, the atoms with lower angular momentum
were selectively removed from the system.

Many exotic features were observed when this technique is applied.
After the atom removal laser was applied continuously, the vortex
number contained in the cloud increased form $180$ to about $230$,
and the vortices began to aggregate to the center of the
condensate, eventually they formed a giant vortex. If the laser
was applied for a short limited time, it was reported in
Ref.\cite{cornell} that a periodic formation and disappearance of
the giant vortex core was observed after applying a strong atom
removal pulse, and in Ref.\cite{cornellsecond} that the Tkachenko
oscillation of the vortex lattice would be found after a weak
pulse. Both kinds oscillation were found damped.
\cite{cornell}\cite{cornellsecond}

A number of questions are raised from these experiments, such as
how to understand the vortex number increase and the vortices
aggregating to the center of condensate, why the strong and weak
laser strength will cause two different kinds of oscillations, how
to understand the relation between the oscillation frequency and
the initial conditions of the condensate, and which mechanism
leads to the damping. We notice that recently there are lots of
theoretical efforts around these experiments, some of them try to
understand these observations numerically\cite{Ballagh}, and some
of them focus on the Tkachenko wave of the vortex lattice
state\cite{Baym}\cite{Bigelow}\cite{stangari}\cite{Salomaa}.
However, in this paper, we will present an explanation to these
questions from another point of view, namely using a dual
$2+1$-dimensional Maxwell electrodynamic (MED) description, and we
will show that the main features observed can be understood
naturally in this framework, although qualitatively in some
aspects.

This dual electromagnetic description of the interacting bosons
model was obtained by the duality transformation early in the
study of Helium superfluidity,\cite{savit}\cite{popov} and then in
the past decades it has been discussed in the contents of
superconductivity and superfluid film by various
authors.\cite{DHlee}\cite{Arovas} Compared to the model discussed
before, we would like to emphasize some important differences of
the system studied here. First, a large amount of angular momentum
have been imparted into the atoms and the rotational invariance is
restored before the formation of condensate, therefore the total
angular momentum of the condensate should be conserved to a
non-zero value. The second, the atom removal laser violates the
current conservation condition. And the third, the system is
dilute and the interaction between atoms is weak, the sound
velocity is consequently relatively small. We will show that these
differences lead to these exotic phenomena in these experiments,
in other words, these observed phenomena reflect some intrinsic
properties of the rotating BEC.

\section{The Dual Picture}

In this section we will firstly make a brief description of the
dual picture for the interacting bosons system, for the details we
refer to the Ref.\cite{popov}\cite{DHlee}\cite{Arovas}. Beginning
with the coherent state path integral formulism,\cite{Negele} we
write down the quantum mechanical propagator
\begin{equation}
Z=\int{D[\psi,\bar{\psi}]\exp\{i\int dt d^2\vec{r} \mathcal{L}\}},
\label{partition}
\end{equation}
where the Lagrangian
\begin{equation}
\mathcal{L}=-\hbar\bar{\psi}\frac{\partial_{t}}{i}\psi-\frac{\hbar^2}{2m}\left|\frac{\vec{\partial}}{i}\psi(r)\right|^2
-\frac{g}{2}\left|\psi(r)\right|^4.  \label{lagrange}
\end{equation}
$g$ is the interaction strength and equals to $4\pi\hbar^2
a_{sc}/m$ for ultracold atomic system, where $a_{sc}$ is the
$s$-wave scattering length between atoms. Here $\psi$ is a complex
field describing the bosonic many-body system, and $\bar{\psi}$ is
its complex conjecture. We can factor $ \psi$ as
$\sqrt{\rho}e^{i\theta}e^{i\theta_{v}}$. The field $\rho$
describes the density of bosons. The phase of $\psi$ contains two
parts, $\theta$ is the smooth part free from singularity and
$\theta_{v}$ is that caused by the presence of vortices.

First of all, we can introduce an auxiliary field $\vec{J}$ by
performing a Hubbard-Stratanovich (H-S) transformation, the
physical meaning of which is the superfluid current. By defining
the three vectors $\tilde{J}\equiv(\hbar\rho,\hbar\vec{J})$ and
$\tilde{\nabla}=(\partial_{t},\vec{\nabla})$, in the low frequency
and long wavelength limits one can neglect the second order
gradient term $|\nabla \rho|^2/\rho$, and obtain that
\begin{eqnarray}
Z=\int D[J]D[\rho] D[\theta] D[\theta_{v}] \exp\left\{i\int dt
d^2\vec{r}
\left(-\tilde{J}\cdot\tilde{\nabla}\theta-\tilde{J}\cdot\tilde{\nabla}\theta_{v}
+\frac{m}{2\rho}|\vec{J}|^2
-\frac{g}{2}\rho^2\right)\right\}.\nonumber\\\label{HStransformation}
\end{eqnarray}
Integrating over $\theta$, one can naturally obtain a
$\delta$-function constraint
$\tilde{\nabla}\cdot\tilde{J}=0$,
which is just the conventional current conservation condition
\begin{equation}
\frac{\partial\rho}{\partial
t}+\vec{\nabla}\cdot\vec{J}=0.\label{currentconservation}
\end{equation}

Owing to this constraint, we can introduce a three-component gauge
field $\tilde{b}$ defined as $(b_{0},\vec{b})$, which satisfies
\begin{equation}
\tilde{J}=\tilde{\nabla}\times\tilde{b},
 \end{equation}
the density and the current of the bosons are thereby determined
by the gauge field. Hence, $\rho=\vec{\nabla}\times\vec{b}$ is
compared to a magnetic field perpendicular to the 2-dimensional
plane. We can expand $\rho$ as $\bar{\rho}+\delta\rho(\vec{r},t)$,
$\bar{\rho}$ is the average density which is compared to an
external magnetic field, while $\delta\rho(\vec{r},t)$ is the
density fluctuation which is compared to the electromagnetic wave.
The electric field $\vec{E}$ is defined as
$\vec{\nabla}b_{0}-\partial_{0}\vec{b}$, and it is perpendicular
to the boson current $\vec{J}$. Therefore the
Eq.(\ref{currentconservation}) can be rewritten as
\begin{equation}
\frac{\partial B}{\partial t}+\vec{\nabla}\times\vec{E}=0,
\end{equation}
which is just the Maxwell equation in a monopole free space-time.

With replacing the $\rho$ by $\bar{\rho}$ in the denominator, the
last two terms in Eq.(\ref{HStransformation})
can also be reexpressed in terms of the gauge field $\tilde{b}$,
which is just the dynamic term $-F^{\mu\nu}F_{\mu\nu}$ of the
gauge field as in the conventional MED. It determines the dynamic
properties of boson density fluctuation, i.e. phonon, and recovers
the well-known result for the sound velocity
$s=\sqrt{g\bar{\rho}/m}$, which plays the same role as the light
velocity in $2+1$-dimensional MED.

The coupling between the vortices and bosons can be illustrated as
\begin{eqnarray}
-\tilde{J}\cdot\tilde{\nabla}\theta_{v}
=-\tilde{b}\cdot\left(\tilde{\nabla}\times\tilde{\nabla}\theta_{v}\right)
=-\tilde{b}\cdot\tilde{J}_{v}.\label{A14}
\end{eqnarray}
Here $\tilde{J}_{v}$ is defined as
$\tilde{\nabla}\times\tilde{\nabla}\theta_{v}$. It can be verified
that the zero-component of $\tilde{J}_{v}$, which is denoted by
$\rho_{v}$, represents the vortex density, and the other two
spatial components represent the vortex current. The
Eq.(\ref{A14}) indicates that the vortices can be viewed as
charged particles, which are coupled to an electromagnetic field
via gauge coupling. The coupling constant, that is the charge $q$
of the charged particles, is unit. In the Coulomb gauge, the
scalar potential $b_{0}$ is instantaneous, one can integrate it
out and obtain the mass of vortex $m_{v}$, which is equal to
${\bar{\rho}\hbar^2}\ln\frac{R}{\Lambda}/(4\pi s^2m)$ for a
uniform condensate, and the mutual logarithmic interaction between
vortices described by
\begin{equation}
\mathcal{L}_{int}=-\frac{\bar{\rho}\hbar^2}{4\pi
m}\int{d^2\vec{r}d^2\vec{r^{\prime}}\rho_{v}(\vec{r})\ln\frac{|\vec{r}-\vec{r^{\prime}}|}{\Lambda}
\rho_{v}(\vec{r^{\prime}})}.\label{log-interaction}
\end{equation}
Here $R$ is the radius of the condensate and $\Lambda$ is the
radius of vortex core.

Thus we have obtained a dual electrodynamic description of
quantized vortices, which is briefly summarized in the following
table. In the dual picture the vortex lattice state corresponds to
the Wigner crystal state of 2-dimensional electrons in the
external magnetic field.
\\

\begin{tabular}{|c|c|}
\hline \bf{Original Picture} & \bf{Dual Picture} \\
\hline vortex & charged particle \\
\hline average & external magnetic field \\ superfluid density
$\bar{\rho}$ &
\\
\hline superfluid current $J$ & electric field $E_{\mu}=\epsilon_{\mu\nu}J_{\nu}$\\
\hline density fluctuation & electromagnetic wave \\
\hline sound velocity & light velocity \\
\hline the coupling between vortex & gauge coupling \\
and sound wave &\\
\hline
\end{tabular}
\\
\\

Such a description can be directly applied to the confined BEC
when it is in the Thomas-Fermi regime. The average density here is
therefore determined by the trapping potential, as well as the
centrifugal repulsive force caused by rotating, and we will take
it as inverted parabola in this regime. Here we would like to
point out that the main assumption leading to the electrodynamic
description is that the vortices are treated as point-like
particles, the physics inside the vortex core is not considered,
as a consequence we take the mass of vortex as its electromagnetic
mass throughout this paper\cite{Arovas}. Besides, in the following
treatment we view the system as quasi-two dimensional and neglect
the density inhomogeneity when estimating the vortex mass. In the
quantum Hall mean field regime\cite{Ho}, it is shown that the
physics of vortex core will become important. However, before
entering this regime, and when the Thomas-Fermi theory is still
valid, these neglected effects will only slightly renormalize the
vortex mass as well as the interaction between vortices.

\section{Explanation of the Experiments}

\subsection{Vortex Number Increasing}
In JILA experiments the rotational invariance is restored before
condensation, therefore the condensate should obey the total
angular momentum conservation constraint. Such a system can be
described by adding a Lagrange multiplier term to the Lagrangian.
\begin{equation}
Z=\int d\psi d\Omega\exp\{i\int d^2
\vec{r}dt\mathcal{L}(\psi,\bar{\psi})-\Omega(\bar{\psi}\hat{L}\psi-L_{0})\}.
\label{langrange-multiply}
\end{equation}
It is obvious that one will obtain a $\delta$-function which means
that the total angular momentum must always equal to $L_{0}$ after
integrating $\Omega$ out. The physical meaning of $\Omega$ is in
fact the effective rotation frequency. Following the procedure
discussed in above section, it is found that only three additional
terms, namely
$-2mb_{0}\Omega+\frac{1}{2}m\Omega^2\rho r^2+\Omega L_{0}$,
will be added to the Lagrangian. One can see that only the zeroth
component of the gauge field responds to the rotation. Integrating
$\Omega$ out and neglecting the density fluctuation, we obtain
\begin{equation}
\mathcal{L}=-\frac{(-2mb_{0}+L_{0})^2}{2m\langle\rho r^2\rangle}+
\textrm{other terms}\label{additional}.
\end{equation}
Following the same step which leads to Eq.(\ref{log-interaction}),
integrating out $b_{0}$, we can obtain a local chemical potential
term in the following Lagrangian
\begin{eqnarray}
\mathcal{L}_{int}=-\int{d^2\vec{r}d^2\vec{r^{\prime}}\rho_{v}(\vec{r})
G(|\vec{r}-\vec{r^{\prime}}|)\rho_{v}(\vec{r^{\prime}})} +\int
d^2\vec{r}\frac{2L_{0}}{\langle\rho r^2\rangle}\left(\int d^2
r^{\prime}G(|\vec{r}-\vec{r^{\prime}}|)\right
)\rho_{v}(r)\nonumber\\ =\mathcal{E}_{int}+\int
d^2\vec{r}\mu(\vec{r})\rho_{v}(\vec{r}).\label{chenmicalporential}
\end{eqnarray}
Here $G(|\vec{r}-\vec{r}^\prime|)$ is the Green's function for the
$b_{0}$ field acquiring a very small mass $2m/\langle\rho
r^2\rangle$, which will vanish in the large atom number limit. The
first term in the Eq. (\ref{chenmicalporential}) indicates that
the vortex system can be viewed as an interacting Coulomb gas
system, and the second term defines a local chemical potential
$\mu(r)$ for vortices emerging from the angular momentum
constraint, which is proportional to $L_{0}/\langle\rho
r^2\rangle$. The atom removal laser, which removes atoms with
lower angular momentum, decreases the atom density $\rho$ with
little change to the total angular momentum $L_{0}$, leads to the
increase of the vortex chemical potential. Notice that the
interaction energy $\mathcal{E}_{int}$ depends quadratically
dependence on vortex density $\rho_{v}$, and the interaction
between vortices are repulsive, so
\begin{equation}
\frac{\partial\mu}{\partial
\rho_{v}}=\frac{\delta^2\mathcal{E}_{int}}{\delta \rho_{v}^2}>0.
\end{equation}
This shows the increase of vortex density.

\begin{figure}[htbp]
\begin{center}
\includegraphics[width=3.3in]
{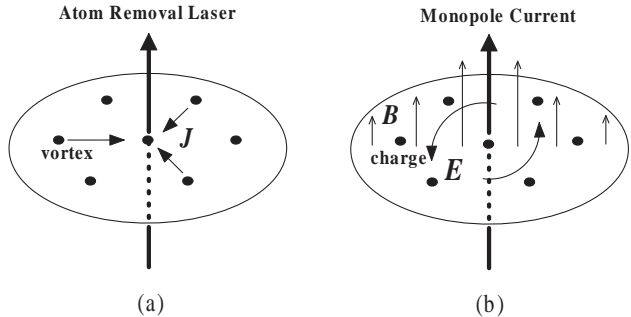} \caption{(a) Atom removal laser induces a current
of atom in the original picture (b) Vortices are viewed as charges
moving under a perpendicular magnetic field in the dual picture.
The thin lines represent the magnetic and electric field, the
thick lines represent the atom removal laser or monopole current.
}
\end{center}
\end{figure}

\subsection{Aggregate} For simplicity we assume that the atom
removal laser acts only at the center point. Extension to the
Gaussian shape laser is straightforward and the qualitative
results will not be changed. When the laser is applied, a
centripetal atom current will be induced and the current
conservation condition will be violated at the center. In the dual
picture, a circular electric field will be generated and the
Maxwell equation should be modified in the following way
\begin{equation}
\frac{\partial B}{\partial
t}+\vec{\nabla}\times\vec{E}=I\delta(\vec{r}).\label{ModifiedME}
\end{equation}
Here the value of $I$ equals to the number of atoms removed from
the condensate per unit time. Therefore it is modelled as a
monopole current running through the plane. The magnitude of the
electric field is approximately $I/(2\pi |\vec{r}|)$ nearby the
center. For the case of continuous laser the vortices as charges
will be accelerated azimuthally by the electric field, and the
Lorentz force will drive the vortices to aggregate toward the
center, and eventually to form a giant vortex.

\subsection{Two Kinds Oscillation} After a pulse laser which only
acts for a short time $\Delta T$, the vortices will acquire a
velocity $\vec{v}$ and leave the balance positions of the lattice.
For the vortices closest to the center, the magnitudes of their
velocities can be approximated by
\begin{equation}
v={E q\Delta T}/{m_{v}}\simeq {I\Delta T\hbar}/(2\pi m_{v} d),
\end{equation}
$d$ is the vortex lattice spacing. Therefore, there are two forces
which can cause the oscillation motion, one is the Lorentz force
$f_{1}$, and the other is the mutual interaction force between
vortices $f_{2}$.  It is easy to obtain the ratio of the two
forces:
\begin{equation}
\frac{f_{1}}{f_{2}}=I\Delta T\frac{m}{m_{v}}.\label{radio}
\end{equation}
For these experiments we can naturally assume that the number of
atoms removed from the condensate $I$ is an increase function of
the power of removal laser. Thus Eq.(\ref{radio}) indicates that
the Lorentz force is dominant for a strong laser pulse, such as
the pulse used in Ref.\cite{cornell}. In this limit, the vortices
will execute circumnutation. When the vortices move very close to
each other, they will merge into a giant vortex, and separate into
several single vortices again when they oscillate away from each
other. The mutual interaction force will be dominant for the weak
pulse case. In this limit, it will cause a collective oscillation
of vortex lattice known as Tkachenko oscillation.

The oscillation frequency $\omega$ depends on the initial
conditions of the condensate, such as the number of atoms and the
rotation frequency $\Omega$. In the strong pulse limit, for the
motion is mostly determined by the Lorentz force,
$\omega={\hbar\rho}/{ m_{v}}$. Because the density $\rho$ is
proportional to the number of atoms $N$ and is a decreasing
function of the rotation frequency $\Omega$, the oscillation
frequency is linearly dependent on $N$ and increases with $\Omega$
decreasing.

\begin{figure}[htbp]
\begin{center}
\includegraphics[width=2.8in]
{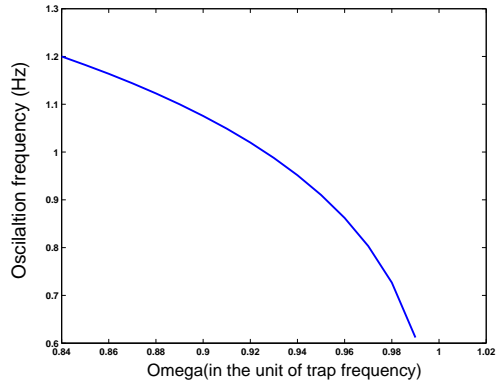} \caption{ The oscillation frequency of vortices
$\omega$ as a function of rotational frequency of condensate
$\Omega$(in the unit of $\omega_{\perp}$). \label{fig:f-Omega}}
\end{center}
\end{figure}

In the weak pulse limit, the oscillation frequency $\omega$ is
proportional to $\sqrt{\rho}/d$.
It is not obvious to tell whether $\omega$ will decrease or not
when the condensate rotates faster, because both $\rho$ and $d$
will decrease with the rotation frequency $\Omega$ increasing. We
make a rough estimate as following: For a fast rotating BEC
confined in a harmonic trap with trapping frequency
$\omega_{\perp}$, the effective confining potential is
$(\omega_{\perp}^2-\Omega^2)r^2/2$. Using the Thomas-Fermi
approximation we can obtain that the condensate density nearby the
center $\bar{\rho}$ is proportional to
$\sqrt{1-\frac{\Omega^2}{\omega_{\perp}^2}}$, and then using the
formula $d^2\propto{1}/{\bar{\rho}_{v}}={\hbar}/(2m\Omega)$, we
can obtain that
\begin{equation}
\omega=C\sqrt{\frac{\Omega}{\omega_{\perp}}\sqrt{1-\frac{\Omega^2}{\omega_{\perp}^2}}}.
\label{fOmega}
\end{equation}
Here $C$ is a complicated constant, which is related to the atom
mass, vortex mass, trap frequency and the density distribution
along the third spatial dimension. The Eq.(\ref{fOmega}) shows
that $\omega$ is a decreasing function of $\Omega$ when
$\Omega/\omega_{\perp}>1/\sqrt{2}$, which is shown in the left
side of Fig.(\ref{fig:f-Omega})\cite{figure}.

From above discussion we can see the atomic current toward the
center is of essential importance for causing oscillations. Hence
if there is any other ways to generate a similar atomic current,
such as using optical dipole force to draw atoms into the middle
of the condensate in the Ref.\cite{cornellsecond}, the similar
oscillation of vortex lattice was observed.

\begin{figure}[htbp]
\begin{center}
\includegraphics[width=2.8in]
{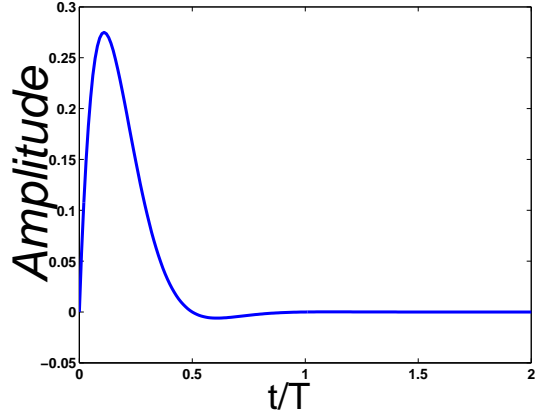} \caption{Heavily damped oscillation amplitude as a
function of time $t$(in the unit of oscillation period $T$)
\label{fig:damping}}
\end{center}
\end{figure}

\subsection{Damping Effect}
One feature of these oscillation motions shown in these
experiments is that the oscillation amplitudes are heavily damped,
especially for the Tkachenko oscillation reported in
Ref.\cite{cornellsecond}. It is quite natural in the view of the
dual picture. Because the coupling between the vortex current and
the phonon field is a gauge coupling, an accelerated vortex will
radiate sound wave in the same way as an accelerated electron
radiating electromagnetic wave. In the view of hydrodynamic
picture, when a vortex is accelerating, the circular current
$\vec{J}$, as well as the singularities of field $\vec{J}$, can
not be stationary in any inertial frame. Due to the current
conservation condition Eq.(\ref{currentconservation}), a time
dependent density fluctuation $\delta\rho(\vec{r},t)$ will be
generated. Hence the oscillating vortices will induce sound wave,
lose their energy and result in the damping of amplitude. In the
dual picture this process can be calculated through a standard MED
calculation.

The retarded Green's function for $2+1 D$ MED is
\begin{equation}
G(\tilde{r},\tilde{r}^\prime)=\frac{\Theta(t-t^\prime)}{2\pi}
\frac{\Theta(s^2(t-t^\prime)^2-(\vec{r}-\vec{r}^\prime)^2)}
{\sqrt{s^2(t-t^\prime)^2-(\vec{r}-\vec{r}^\prime)^2}}.
\end{equation}
And the gauge potential $\tilde{b}_{rad}$ for the radiation field
is
\begin{equation}
\tilde{b}_{rad}=\int\limits_{-\infty}^{t-|\vec{r}-\vec{r}(t^\prime)|/s}
dt^\prime\frac{\tilde{V}(t^\prime)}{\sqrt{s^2(t-t^\prime)^2-(\vec{r}-\vec{r}(t^\prime))^2}}.
\end{equation}
Here $\tilde{V}(t^{\prime})$ is the vortex $3$-current. Unlike the
$3+1$D case, the Green's function contains the step function
$\Theta$ instead of Dirac's $\delta$ function, so one can not
easily find a general expression for the radiation energy of a
charge with arbitrary acceleration.\cite{PAo}\cite{MaxwellE} Here
we only consider two special cases, the linear oscillation
$\vec{V}(t^\prime)=(v\cos(\omega t^\prime), 0)$, and circular
motion $\vec{V}(t^\prime)=(-v\sin(\omega t^\prime),v\cos(\omega
t^\prime))$. Following the approximation made in Ref.\cite{PAo},
we replace $|\vec{r}-\vec{r}(t^\prime)|$ with $R$ and calculate
the energy radiation along large circle $|r|=R$. Then we can
perform an analytical calculation for the average energy radiated
over one period,
\begin{equation}
P=\frac{1}{T}\int\limits_{0}^{T}\int\limits_{0}^{2\pi}dtd\theta
r\vec{\sigma}\cdot\hat{r},
\end{equation}
where $\vec{\sigma}$ is the Poynting vector. For these two special
cases, the result is
\begin{equation}
P=\frac{\rho\hbar^2\omega v^2}{2m s^2}=\frac{\hbar^2\omega
v^2}{2g}.\label{radiationpower}
\end{equation}
The radiation rate is inverse proportional to the interaction
strength $g$, so the damping effect is remarkable for the weak
interacting boson gases. Recalling the approximate expression for
the vortex mass $m_{v}=\bar{\rho}\hbar^2\ln(R/\Lambda)/(4\pi
s^2m)$, Eq.(\ref{radiationpower}) can be reexpressed as
\begin{eqnarray}
P=\frac{1}{2}\frac{\rho\hbar^2}{s^2m}\omega v^2\nonumber=
\frac{1}{2}\frac{4\pi\omega}{\ln\frac{R}{\Lambda}}m_{v}v^2=\frac{4\pi\omega}{\ln\frac{R}{\Lambda}}E.
\end{eqnarray}
Hence
\begin{equation}
\frac{dE}{dt}=-\frac{4\pi\omega}{\ln\frac{R}{\Lambda}}E,
\end{equation}
this indicates that the oscillation energy exponentially decays as
$\exp\{-t/\tau\}$, and the ratio of the exponential damping time
$\tau$ to the oscillation period $T$ is equal to
$\ln\frac{R}{\Lambda}/(8\pi^2)$. A typical oscillation curve in
the current experiment conditions is schematically shown in the
right side of Fig.\ref{fig:damping}, one can see that the
oscillation amplitude will decay rapidly within one period, and
this is qualitatively coincident with the experimental
observation\cite{cornell}\cite{cornellsecond}. However, there are
also a little quantitatively disagreement between the theoretical
damping curve and the experiments, in Ref.\cite{cornell} the
amplitude vanishes within about three periods and in
Ref.\cite{cornellsecond} it is about one and a half periods. This
disagreement occurs because in the real case the vortex mass
should be modified due to the spatial density inhomogeneity, the
third spatial dimension effect and the edge effect. The mass of
vortex, which is proportional to the self-energy, is therefore
changed with the position of vortex.\cite{off-vortex} It is
expected that the damping curve will agree better with the
experiments when these subtle effects are carefully considered.

\section{Summary}

According to the effective Maxwell electrodynamic model the
vortices in the BEC can be viewed as an ensemble of classical
charges moving in an external magnetic field. So far we have
applied this description to understand the observed phenomena of
vortices induced by selective removal atoms from a fast rotating
BEC. The atom removal laser applied to the center of the
condensate will cause two leading effects in the view of dual
picture. One is the increase of the vortex chemical potential and
then subsequently the increase of vortex density. The other is
that it will induce a circular electric field, and each vortex
will acquire an azimuthal velocity. We clarified that there are
two forces that can cause oscillation motion, and discussed the
the relation between the oscillation frequencies and the number of
atoms together with the effective rotation frequency. We showed
that the dispersion mechanism for the vortices oscillation comes
from the emission of phonon, which is the same as the radiation of
the electromagnetic wave of an accelerated charge. And the damping
effect is more remarkable in these experiments because the
interaction strength is weak.

In this paper the dual description of vortices has been used to
explain current experiments. Although it is not yet a detailed
quantitative description, it provides a natural physics picture of
most key experimental observations, which convinces us to believe
that this description is valid to the extent that the BEC is fast
rotated but still not entering the quantum Hall mean field regime.
We hope this will start a new route to study the issues in the
rapidly rotating BEC system.

\textit{Acknowledgements}: HZ would like to thank Professor C. N.
Yang for encouragement. And the authors would like to acknowledge
Z.Y. Weng, T.L. Ho, L. Chang, R. L$\ddot{u}$ and X.L. Qi for
helpful discussions. This work is supported by National Natural
Science Foundation of China ( Grant No. 10247002 )



\begin{thebibliography}{99}
\bibitem{figure}
We have made a rough estimate that the frequency given by
equation(\ref{fOmega}) is the the same order of the experiment
date. And we use the experiment date $\Omega/\omega_{\perp}=0.84$,
$\omega=1.2 Hz$ reported in Ref.\cite{cornellsecond} to determine
the constant $C$
\end{thebibliography}
\end{document}